\documentclass{article}
\newcommand{\be}{\begin{equation}}
\newcommand{\ee}{\end{equation}}

\def\simlt{\stackrel{<}{{}_\sim}}

\def\bea{\begin{eqnarray}}
\def\eea{\end{eqnarray}}
\def\bean{\begin{eqnarray*}}
\def\eean{\end{eqnarray*}}

\newcommand{\barr}{\begin{array}}
\newcommand{\earr}{\end{array}}

\newcommand{\bed}{\begin{displaymath}}
\newcommand{\eed}{\end{displaymath}}
\newcommand{\bal}{\begin{array}{ll}}
\newcommand{\eal}{\end{array}}

\def\bvec#1{\raise1.5ex\hbox{$\rightarrow$}\mkern-16.5mu #1}

\begin{document}

\pagestyle{empty}

\rightline{UFIFT-HEP-05-6}
\rightline{MCTP-05-65}
\vspace*{1.5cm}

\begin{center}

\LARGE{Cabibbo Haze in Lepton Mixing\\[20mm]}

\large{AseshKrishna Datta$^{1,2}$,
Lisa Everett$^{1}$,
and
Pierre Ramond$^{1}$
\\[8mm]}

\it{1. Institute for Fundamental Theory, 
Department of Physics\\ 
University of Florida, Gainesville, FL, 32611, USA\\[5mm]}
\it{2. Michigan Center for Theoretical Physics, Department of 
Physics\\ University of Michigan, Ann Arbor, MI, 48109, 
USA\\[15mm]}
 
\large{\rm{Abstract}} \\[7mm]

\end{center}

\begin{center}

\begin{minipage}[h]{14cm}
Quark-lepton unification suggests Cabibbo-sized effects in 
lepton mixings, which we call Cabibbo haze.  We give simple examples and 
explore possible Wolfenstein-like parametrizations of the MNSP matrix 
which incorporate leptonic Cabibbo shifts.  We find that the size of the 
CHOOZ angle is not always correlated with the observability of CP 
violation.
\end{minipage}

\end{center}

\newpage

\pagestyle{plain}

\section{Introduction}
With recent experimental progress in the neutrino sector, a quantitative 
picture of the Maki-Nakagawa-Sakata-Pontecorvo (MNSP) lepton mixing 
matrix \cite{Maki:1962mu,Pontecorvo:1957cp} has begun to emerge.   In 
contrast to the Cabibbo-Kobayashi-Maskawa (CKM) quark mixing matrix, 
which is approximately the identity matrix up to effects of order the 
Cabibbo angle $\lambda\equiv \sin\theta_c\sim 0.22$,
the MNSP matrix contains two large angles 
$\theta_{\odot}$, $\theta_{\oplus}$, and one small angle $\theta_{13}$: 
\be \mathcal{U}^{}_{MNSP}~\simeq~\pmatrix{\cos\theta_{\odot}& 
\sin\theta_{\odot}&\sin\theta_{13} \cr
-\cos\theta_{\oplus}~\sin\theta_{\odot}&\cos\theta_{\oplus}~ 
\cos\theta_{\odot}&\sin\theta_{\oplus} 
\cr 
\sin\theta_{\oplus}~\sin\theta_{\odot}&-\sin\theta_{\oplus}~ 
\cos\theta_{\odot}&\cos\theta_{\oplus}}\ ,
\ee
for which combined data from the solar (with SNO-salt \cite{salt} and
Super-Kamiokande (SK) \cite{sksolar}),
atmospheric (from the latest SK data \cite{skatm}),
reactor (KamLAND \cite{kamland}, 
CHOOZ and Palo Verde \cite{chooz}), and accelerator experiments (K2K) 
\cite{k2k} 
indicates at
3$\sigma$ the following
constraints on the atmospheric and CHOOZ angles \cite{global}:\footnote{Significant 
improvements in constraining $\theta_{13}$ (down to $3^\circ$ or below) 
are foreseen at the planned reactor neutrino experiments, superbeams 
and/or neutrino factories \cite{nufact}; not much improvement in the 
atmospheric sector is foreseen \cite{kajita}.} 
\be
\quad \theta^{}_\oplus~=~ {45^\circ_{}}^{\,+10^\circ}_{\,-10^\circ} \
;\quad \theta_{13}< ~13^\circ_{},
\label{data1}
\ee
while the solar angle is 
constrained to a rather precise range of 
\be
\theta_{\odot}={32.5^\circ}^{+2.4^\circ}_{-2.3^\circ},
\label{data2}
\ee
as indicated by the SNO-salt data when combined with the data from 
the KamLAND reactor experiment.
Note that the limiting value of $\theta_{13}$ is approximately equal to 
$\theta_c$. 

Understanding the origin of the MNSP mixing angles provides a new 
perspective and challenges for addressing the flavor 
puzzles of the Standard Model (SM). In this letter, we work in the 
context of quark-lepton grand unification \cite{patisalam}, for which all 
available quark and lepton sector data can be synthesized in 
the quest toward a credible theory of flavor.  
The qualitative differences between the quark and lepton mixing
matrices, while perhaps surprising, are not inconsistent
with grand unification.  The seesaw mechanism \cite{SEESAW}
introduces a new unitary matrix into the MNSP matrix, with no 
analogue in the quark sector \cite{datta}, that can provide a source for 
the discrepancy. Explaining the observed pattern of two large 
and one small lepton mixing angles without fine-tuning in a three-family 
model is the Achilles' heel for most model-building attempts.

Wolfenstein's parametrization of the CKM matrix \cite{wolfenstein} as an 
expansion in powers of $\lambda$ not only has well-known practical 
advantages for phenomenology, but also provides a theoretical framework 
for examining the quark sector in the $\lambda \rightarrow 0$ limit.
One aim of this letter is to seek a similar parametrization of the MNSP 
matrix.  We begin by demonstrating that grand unification implies that 
Cabibbo-sized effects are expected in the lepton as well as the quark 
sector. Such effects create a {\it Cabibbo haze} in lepton flavor mixing, 
keeping in mind that unlike the CKM matrix, the MNSP 
matrix is not known in the $\lambda\rightarrow 0$ limit, except that 
it contains two large angles.  Leptonic Cabibbo-sized perturbations can 
then shift the
atmospheric and solar angles from their unknown initial values by $\sim
\theta_c$, provide the dominant source for the CHOOZ angle, and dictate
the size of CP-violating effects. Although the data is not yet precise
enough to pinpoint a particular Wolfenstein-like parametrization, 
regularities may emerge upon more precise measurements of the MNSP 
parameters.

\section{Seesaw and Grand Unification}
In the SM, $\Delta I_{\rm w}=1/2$ electroweak breaking generates 
Dirac masses for the quarks and charged leptons through Yukawa couplings.  
These Dirac mass matrices are diagonalized by biunitary transformations, in 
which ${\cal U}^{}_{2/3}$,  ${\cal U}^{}_{-1/3}$, and ${\cal U}^{}_{-1}$ 
are the rotation matrices for the left-handed states, while ${\cal 
V}^{}_{2/3}$,  ${\cal V}^{}_{-1/3}$, and ${\cal V}^{}_{-1}$ are the 
rotation 
matrices for the right-handed states.  Physical observables depend only 
on the CKM matrix
\be
{\cal U}^{}_{CKM}~=~{\cal U}^{\dagger}_{2/3}\,{\cal U}^{}_{-1/3}. 
\ee
In the original formulation of the SM, all three neutrinos are massless 
and lepton mixing is unobservable. Nonvanishing neutrino masses require 
one to go beyond the SM, and add {\it e.g.} one right-handed neutrino 
for each family.  Electroweak breaking then also generates Dirac mass 
terms for the neutrinos, with 
${\cal M}^{(0)}_{Dirac}={\cal U}^{}_{0}{\cal D}_0{\cal 
V}^{\dagger}_{0}$.
If there is no seesaw, the Dirac mass eigenvalues $m^{{\rm D}}_{i}$ 
(${\cal D}_0={\rm diag}(m^{{\rm D}}_{i})$) are the physical neutrino masses.  
However, large $\Delta I_{\rm w}^{}=0$ mass terms for the right-handed 
neutrinos arise naturally, as they are unsuppressed by gauge quantum 
numbers. This Majorana mass matrix ${\cal 
M}^{(0)}_{Maj}$ has entries which can be much larger than the 
electroweak scale.  After the seesaw, 
\be
{\cal M}^{(0)}_{Seesaw}~=~{\cal M}^{(0)}_{Dirac}\,
\frac{1}{{\cal M}^{(0)}_{Maj}}\,{\cal M}^{(0)\,T}_{Dirac}\ ,
\ee
which can be reexpressed \cite{datta} as
\bea{\cal M}^{(0)}_{Seesaw}&=&{\cal U}^{}_{0}\,
{\cal D}_0^{}\,{\cal V}^{\dagger}_{0}\,\frac{1}{{\cal M}^{(0)}_{Maj}}\,
{\cal V}^{*}_{0}\,{\cal D}_0^{}\,{\cal U}^{T}_{0}\ ,\cr &=&{\cal U}^{}_{0}\,
{\cal C}\,{\cal U}^{T}_{0}\ .\eea
$\cal C$ is the central matrix
\be{\cal C}~\equiv~{\cal D}_0^{}\,{\cal V}^{\dagger}_{0}\,
\frac{1}{{\cal M}^{(0)}_{Maj}}\,{\cal V}^{*}_{0}\,{\cal D}_0^{}\ ,\ee
diagonalized by the unitary matrix ${\cal F}$ 
\be{\cal C}~=~{\cal F}\,{\cal D}^{}_\nu\,{\cal F}^T_{}\ ,\ee 
where ${\cal D}_\nu$ is the diagonal matrix of the physical neutrino 
masses $m_1$, $m_2$, and $m_3$.
The MNSP matrix 
can then immediately be written in the 
suggestive form \be
\label{ufumnsp}
{\cal U}^{}_{MNSP}~=~ {\cal U}^{\dagger}_{-1}\,{\cal 
U}^{}_{0}\,{\cal  F}\ .\ee 
Eq.~(\ref{ufumnsp}) highlights the differences between ${\cal 
U}^{}_{MNSP}$ and ${\cal U}^{}_{CKM}$, and provides the basis for our 
discussion.

Grand unification suggests connections between the MNSP and CKM matrices.  
The simplest Higgs structures lead to the following relations 
\be SU(5)~~~:~~~~~~~~{\cal M}^{(-1/3)}_{}~\sim~{\cal M}^{(-1)\,T}_{}\ .\ee
\be SO(10):~~~~~~~~{\cal M}^{(2/3)}_{}~\sim~{\cal M}^{(0)}_{Dirac}\ ,\ee
which imply
$${\mathcal U}^{}_{-1/3}~\sim~{\mathcal V}^{*}_{-1}\ ;\qquad {\mathcal
U}^{}_{2/3}~\sim~{\mathcal U}^{}_{0},\ $$ 
so that
\be
{\mathcal U}^{}_{MNSP}= {\mathcal V}^T_{-1/3}\,{\mathcal 
U}^{}_{-1/3}\,{\mathcal U}^{\dagger}_{CKM}\,{\mathcal F}\ .\ee
Models can then be categorized according to the structure of the  
charge  $-1/3$ Yukawa couplings and the number of large 
angles in $\mathcal F$.  A particularly 
illustrative example \cite{ramond} is the class of models with symmetric 
${\mathcal M}^{}_{-1/3}$, for which 
$${\mathcal U}^{}_{-1/3}~=~{\mathcal V}^*_{-1/3}\ ,$$
which implies that the MNSP and CKM matrices are simply related
\be {\mathcal U}^{}_{MNSP}~=~ {\mathcal U}^{\dagger}_{CKM}\,\,{\mathcal 
F}\
\label{udagf}.
\ee
In this case, $\mathcal F$ must contain two large mixing angles 
$\eta_{\odot}$ and $\eta_{\oplus}$.  As we will discuss in the next 
section, Eq.~(\ref{udagf}) then implies that 
the solar angle 
$\eta_{\odot}$ experiences a 
Cabibbo shift $\theta_{\odot}\sim \eta_{\odot} \pm \lambda 
\cos\eta_{\oplus}\sim\eta_{\odot} 
\pm  \lambda/\sqrt{2}$,
and $\theta_{13}\sim {\lambda}\sin\eta_{\oplus} 
\sim \lambda/\sqrt{2}$ due to the ${\cal O}(\lambda)$ $1-2$ mixing in 
${\cal U}_{CKM}$.

The above class of models provides well-motivated examples of 
leptonic Cabibbo shifts, but it is by no means the only theoretical 
possibility.  In the context of grand unification, the data also can hint 
that the mixing matrix is initially 
bimaximal ($\eta_{\odot}=\eta_{\oplus}=45^{\circ}$), with the solar angle 
shifted by a full-strength Cabibbo shift: $\theta_{\odot}\sim 
\eta_{\odot} - \theta_c$ 
\cite{bimax,Giunti:2002ye,Rodejohann,Cheung:2005gq}.
While these examples can be motivated by flavor theories, one should keep 
in mind that the data is not yet precise enough to 
select particular scenarios and the values of the large angles are not 
known in the $\lambda\rightarrow 0$ limit (if indeed that limit is 
meaningful for theory).  Hence, we will now explore parametrizations of 
the MNSP matrix which incorporate such leptonic Cabibbo effects purely 
from 
a phenomenological standpoint.

\section{Systematics}
The Wolfenstein parametrization of the CKM matrix is based on the idea 
that the observed hierarchical quark mixing angles can be understood as 
powers of the Cabibbo angle $\lambda$, with ${\mathcal U}_{CKM}=1$ in the  
$\lambda \rightarrow 0$ limit: 
\be
\mathcal{U}_{CKM}=1+\mathcal{O}(\lambda).
\ee
The mixing angles of the quark sector (which have been very precisely 
determined using direct measurements and unitarity constraints) have the 
following hierarchical structure \cite{wolfenstein}:  
\be
{\cal 
U}_{CKM}=\pmatrix{1-\frac{\lambda^2}{2}&\lambda&A\lambda^3(\rho-i\eta) \cr 
-\lambda &1-\frac{\lambda^2}{2}&A\lambda^2 \cr A\lambda^3(1-\rho-i\eta) 
&-A\lambda^2&1}+\mathcal{O}(\lambda^4).
\ee
In the above, $\lambda$ and $A$ are well known ($\lambda=0.22$, 
$A\simeq0.85$), but $\rho$ and $\eta$ are less precisely determined 
\cite{pdg}.  The CP-violating phase 
at lowest order is in the most hierarchically suppressed $1-3$ mixing, but 
any other choice leads to the same rephasing invariant measure of CP 
violation.  As measured by the JDGW invariant \cite{jarlskog,greenberg},
CP violation in the quark sector is small
\be
J^{(CKM)}_{CP}\simeq A^2\lambda^6\eta,
\ee
due to the small mixing angles, not the CP-violating phase, which  
is the only (possibly) large angle in ${\cal U}_{CKM}$.

To seek a similar parametrization for the MNSP matrix, we assume a 
$\lambda$ expansion of the form  
\be
\mathcal{U}_{MNSP}=\mathcal{W}+\mathcal{O}(\lambda),
\ee 
in which the starting matrix ${\cal W}$ has two large angles 
$\eta_{\oplus}$ and $\eta_{\odot}$, corresponding to the ``bare" values of 
the atmospheric and solar angles, and a vanishing CHOOZ angle: 
\be
{\cal W}={\cal R}_1(\eta_{\oplus}){\cal 
R}_{3}(\eta_{\odot})\equiv \pmatrix{1&0&0\cr 0&\cos 
\eta_{\oplus}&\sin\eta_{\oplus} \cr 
0&-\sin\eta_{\oplus} &\cos\eta_{\oplus}}
\,\pmatrix{\cos\eta_{\odot} &\sin\eta_{\odot}&0\cr -\sin\eta_{\odot}&\cos 
\eta_{\odot}&0\cr
0&0&1}.
\label{wdef}
\ee
${\cal W}$ is then perturbed by a unitary matrix ${\cal V(\lambda)}$. 
Unlike the case of the quarks, the perturbation matrix does not 
generically commute with the starting matrix
\be 
[{\mathcal 
W}\,,\,{{\mathcal 
V}(\lambda)}\,]~\neq~0\ ,
\ee
resulting in three possible types of Cabibbo shifts (see also 
\cite{ramond,Giunti:2002ye,Rodejohann,Cheung:2005gq}):
\begin{itemize}
\item {\it Right Cabibbo Shifts}: 
\be
{\cal U}_{MNSP}={\cal W}\,{\cal V}(\lambda)
\ee 

\item {\it Left Cabibbo Shifts}: 
\be
{\cal U}_{MNSP}={\cal V}(\lambda)\,{\cal W}
\ee  
\item {\it Middle Cabibbo Shifts}: 
\be
{\cal U}_{MNSP}={\cal R}_{1}(\eta_{\oplus})\,{\cal V(\lambda)}\,{\cal 
R}_{3}(\eta_{\odot})
\ee  
\end{itemize}
Given $\eta_{\odot}$ and $\eta_{\oplus}$, one can 
choose ${\cal V}(\lambda)$ to shift the atmospheric and solar 
angles into the range allowed by the data, and study the resulting 
implications for $\theta_{13}$ in the three scenarios.  
A novel feature of our parametrizations is that the JDGW invariant 
depends not only on the type of Cabibbo shift, but also on how the 
CP-violating phase is introduced in ${\cal V}(\lambda)$.  
$J^{(MNSP)}_{CP}$ is typically expected to be much larger 
than $J^{(CKM)}_{CP}$ due to the larger mixing angles, and 
can be as large as $\sim \lambda \sin\delta$ 
(for $\theta_{13}\sim \theta_c$ and $\delta\sim {\cal O}(1)$).  

To illustrate these ideas, let us consider models for 
which Eq.~(\ref{udagf}) holds, which are left Cabibbo shifts with 
${\cal F}$ given by Eq.~(\ref{wdef}) and ${\cal V}(\lambda)={\cal 
U}^{\dagger}_{CKM}$. The shifts in the angles 
are given to ${\cal 
O}(\lambda^2)$ by
\begin{eqnarray}
\theta_{\oplus} & = & \eta_{\oplus}-(A+\frac{1}{4}\sin 2\eta_{\oplus}) 
\lambda^2 , \\
\theta_{\odot} & = &\eta_{\odot}-\cos\eta_{\oplus}\lambda , \\
\theta_{13} & = & -\lambda\sin\eta_{\oplus}.\end{eqnarray}
Using the data to constrain $\mathcal{F}$, assuming the central values 
$\theta_{\oplus}=45^{\circ}$ and $\theta_{\odot}=32.5^{\circ}$ for 
simplicity,  $\eta_{\oplus}\simeq48^{\circ}$ (the shift of $\sim3^{\circ}$ 
is a typical  ${\cal O}(\lambda^2)$ correction), 
$\eta_{\odot}\simeq41^{\circ}$, and $\theta_{13}=\sin^{-1}(- 
\lambda\sin\eta_{\oplus})\simeq9.4^{\circ}$.  
The JDGW invariant is given by 
\be
\label{ufj}
J^{(MNSP)}_{CP}=\mp \frac{1}{4}\cos\eta_{\oplus}\sin 2\eta_{\oplus}\sin 
2\eta_{\odot} A\lambda^3\eta
+{\cal O}(\lambda^4),
\ee
which is $\sim 10^{-3}$ (setting the CKM 
parameter $\eta \sim 0.4$). As the MNSP matrix has two ${\cal O}(1)$ and 
one ${\cal O}(\lambda)$ mixing angles, the effective MNSP phase is 
suppressed by ${\cal O}(\lambda^2)$, even though 
$J^{(MNSP)}_{CP}$ is larger than $J^{(CKM)}_{CP}$.

\section{Parametrizations}
We now analyze Wolfenstein-like parametrizations of the MNSP matrix which 
incorporate the different types of leptonic Cabibbo 
shifts.  As Cabibbo 
shifts are at most $\sim \theta_c$ (for ${\cal O}(\lambda)$ 
perturbations), the bare angles $\eta_{\odot}$ 
and $\eta_{\oplus}$ can {\it e.g.} be in the approximate ranges $15^\circ< 
\eta^{}_\odot< 45^\circ$, $30^\circ< \eta_\oplus< 
60^\circ $.  One should also keep in mind that the error bars on 
$\theta_{\oplus}$ and 
the bound on $\theta_{13}$ are roughly ${\cal O}(\lambda)$, while the 
error bars on $\theta_{\odot}$ 
are of ${\cal O}(\lambda^2)$.

Given $\eta_{\odot}$ and $\eta_{\oplus}$, the perturbation matrix ${\cal 
V}(\lambda)$ must shift the two large angles in the range allowed by the 
data.
 Whether the 
CHOOZ angle is shifted by $\sim \theta_c$ or a subleading contribution 
depends on the details of the ${\cal O}(\lambda)$ perturbations 
in ${\cal V}(\lambda)$ and the type of Cabibbo shift, leading to three 
basic categories: 
\begin{itemize}
\item {\it Single shift} models. These models have only one  ${\cal 
O}(\lambda)$ perturbation in ${\cal V}(\lambda)$, which can be either in 
the $1-2$ mixing ${\cal 
V}_{12} \sim \lambda$, the $2-3$ mixing ${\cal V}_{23}\sim \lambda$, 
or the $1-3$ mixing ${\cal V}_{13}\sim \lambda$.

\item {\it Double shift} models.  These models have two ${\cal 
O}(\lambda)$ perturbations.  There are three possibilities: ${\cal
V}_{12} \sim {\cal V}_{23} \sim \lambda$, ${\cal 
V}_{12}\sim {\cal V}_{13} \sim \lambda$,
or ${\cal V}_{13}\sim {\cal V}_{23}\sim \lambda$. 

\item {\it Triple shift} models. These models have  ${\cal
V}_{12} \sim {\cal V}_{23} \sim {\cal V}_{13} \sim \lambda$.
\end{itemize}
Focusing for the moment on single shifts, there are several 
broad classes of parametrizations:\\

\noindent $\bullet$ {\it Cabibbo-shifted atmospheric 
angle:}\\
\noindent In this class of models, the perturbation ${\cal V}(\lambda)$ 
shifts 
the atmospheric angle by $\sim \theta_c$.  One possibility is that 
\be
\label{23pert}
{\cal V}(\lambda)=\pmatrix{1&0&0\cr
0& 1 & a\lambda \cr
0& -a\lambda&1}+{\cal O}(\lambda^2),
\ee
with $a\sim {\cal O}(1)$.  The solar angle remains unshifted 
($\theta_{\odot}=\eta_{\odot}$).  The shifts in the atmospheric 
and CHOOZ angles depend on the type of shift scenario.  Right shifts 
\be
{\cal U}_{MNSP}=
\label{rcs23}
\mathcal{R}_1(\eta_{\oplus})
\mathcal{R}_3(\eta_{\odot})
\pmatrix{1&0&0\cr 0& 1 & a\lambda \cr 0& -a\lambda&1} 
\ee
lead to 
\bea
\label{atmrshift}
\theta_{\oplus}&\simeq&\eta_{\oplus}+a\lambda \cos\eta_{\odot} , \nonumber 
\\
\theta_{13}&\simeq&a \lambda \sin\eta_{\odot},
\eea
Due to the form of Eq.~(\ref{rcs23}), the size of Cabibbo shifts are 
$\eta_{\odot}$-dependent.  Left shifts 
\be
{\cal U}_{MNSP}=\pmatrix{1&0&0\cr
0& 1 & a\lambda \cr
0& -a\lambda&1}
\mathcal{R}_1(\eta_{\oplus})
\mathcal{R}_3(\eta_{\odot})
\ee
and middle shifts
\be
{\cal U}_{MNSP}=\mathcal{R}_1(\eta_{\oplus})
\pmatrix{1&0&0\cr 0& 1 & a\lambda \cr 0& -a\lambda&1}
\mathcal{R}_3(\eta_{\odot})\ee
give the same results (to this order in $\lambda$):
\bea
\theta_{\oplus}&\simeq&\eta_{\oplus}+a\lambda , \nonumber\\ 
\theta_{13}&\simeq&0.
\eea
The atmospheric angle is now shifted by a full-strength Cabibbo effect, 
and the CHOOZ angle is a higher order effect.  If, however, 
\be
{\cal V}(\lambda)=\pmatrix{1&0&a \lambda\cr
0& 1 & 0 \cr
-a\lambda&0&1}+{\cal O}(\lambda^2),
\label{13pert}
\ee
right shifts lead to 
\bea
\theta_{\oplus}&\simeq&\eta_{\oplus}-a\lambda \sin\eta_{\odot} 
, \nonumber\\ 
\theta_{13}&\simeq&a \lambda \cos\eta_{\odot},
\eea
while the left and middle shifts leave the atmospheric angle 
unchanged at this order.\\

\noindent $\bullet$ {\it Cabibbo-shifted solar angle}: \\
\noindent In this case, ${\cal V}(\lambda)$ shifts the 
solar angle by $\sim \theta_c$.  One possibility is that 
\be
\label{12pert}
{\cal V}(\lambda)=\pmatrix{1&a \lambda&0\cr
-a\lambda& 1 & 0 \cr
0&0&1}+{\cal O}(\lambda^2),
\ee
just as in the CKM. The atmospheric angle is unchanged to 
this order in $\lambda$ for all scenarios.  
For right shifts
\be
{\cal U}_{MNSP}=
\label{rcs12}
\mathcal{R}_1(\eta_{\oplus})
\mathcal{R}_3(\eta_{\odot})
\pmatrix{1&a\lambda&0\cr
-a\lambda& 1 & 0 \cr
0& 0&1} 
\ee
and middle shifts
\be
{\cal U}_{MNSP}=
\mathcal{R}_1(\eta_{\oplus})
\pmatrix{1&a\lambda 
&0\cr -a\lambda 
& 1 & 0
\cr 0& 0&1}
\mathcal{R}_3(\eta_{\odot})
\ee
the leading order shifts in the solar and CHOOZ angles are
\bea
\theta_{\odot}&\simeq&
\eta_{\odot}+a\lambda , \nonumber\\ 
\theta_{13}&\simeq&0.
\eea
In contrast, left shifts 
\be
{\cal U}_{MNSP}=\pmatrix{1&a\lambda&0\cr
-a\lambda& 1 & 0 \cr
0& 0&1}
\mathcal{R}_1(\eta_{\oplus})
\mathcal{R}_3(\eta_{\odot})
\ee
yield
\bea
\theta_{\odot}&\simeq&
\eta_{\odot}+a\lambda\cos\eta_{\oplus}, \nonumber\\
\theta_{13}&\simeq&a\lambda\sin\eta_{\oplus}.
\eea
Left shifts also works if ${\cal V}(\lambda)$ is 
given by Eq.~(\ref{13pert}), in which case
\bea
\theta_{\odot}&\simeq&
\eta_{\odot}-a\lambda , \nonumber\\ 
\theta_{13}&\simeq&a\lambda\cos\eta_{\oplus}.
\eea
Note that here the Cabibbo shifts are sized by 
$\eta_{\oplus}$-dependent factors.\\

\noindent $\bullet$ {\it Unshifted solar and 
atmospheric angles}:\\
\noindent In this case, the starting values of the large angles are very 
close to their central values.  
There are two possibilities: ${\cal 
V}(\lambda)$ has no entries linear in $\lambda$, or 
${\cal V}(\lambda)$ is given by Eq.~(\ref{13pert}),  in which 
case middle Cabibbo shifts 
\be
{\cal U}_{MNSP}=
\mathcal{R}_1(\eta_{\oplus})\pmatrix{1&0&a 
\lambda \cr
0& 1 & 0 \cr
-a\lambda&0&1}
\mathcal{R}_3(\eta_{\odot})
\ee
lead to $\theta_{13}=a\lambda$ and unshifted large angles.\\

\noindent 
A similar analysis can be carried out for the double and triple Cabibbo 
shift scenarios.  In analogy with the single shift models, for which 
either one or two of the three mixing angles receive an ${\cal 
O}(\lambda)$ shift,  either two or all three of the mixing angles can 
receive large Cabibbo shifts in double shift models.  Triple shifts 
lead to ${\cal O}(\lambda)$ shifts in all three mixing angles. 
In Table~\ref{doubletripleshifts}, results are 
shown for double and triple shifts which are parametrized using one ${\cal 
V}(\lambda)$ matrix (with two or three ${\cal O}(\lambda)$ entries) 
and right, left, or middle shifts.  

We do not discuss more complicated ways to introduce ${\cal 
O}(\lambda)$ effects in detail, 
as they add little to our qualitative conclusions.  For example, double 
and triple shift models can be constructed using two or three 
single shifts (with ${\cal V}(\lambda)$ given by Eq.~(\ref{23pert}), 
Eq.~(\ref{13pert}), or Eq.~(\ref{12pert})) and combinations of 
right, left, and middle shifts. Triple shift models can also incorporate 
combinations of double and single shifts.  These models lead to 
shifts with similar order of Cabibbo suppression,
though their sizes can be different.  An illustrative example 
is a double shift 
obtained by a left shift with ${\cal V}_{12}\sim a \lambda$ and a middle 
shift with ${\cal V}_{13}\sim a^\prime \lambda$ 
($a,\ a^\prime\sim{\cal O}(1)$), which yields  
$\theta_{\odot}=\eta_{\odot}+a\cos\eta_{\oplus}\lambda$, 
$\theta_{\oplus}=\eta_{\oplus}$, 
and $\theta_{13}=(a^\prime+a\sin\eta_{\oplus})\lambda$.  

The various parametrizations can also be classified in terms of 
the predictions for $\theta_{13}$. Perturbations with ${\cal V}_{13}\sim {\cal 
O}(\lambda)$ (such as in Eq.~(\ref{13pert})) always lead to $\theta_{13}$ 
of order $\theta_c$. It is also possible to  obtain a shift in 
$\theta_{13}$ of that size through ${\cal O}(\lambda)$  entries in $1-2$ 
mixing with left shifts and $2-3$ mixing with right shifts.  

Finally, we note that based on this leading order analysis, many models 
can be constructed by specifying the bare angles $\eta_{\odot},\ 
\eta_{\oplus}$ and including subleading perturbations.  
We choose not to do this at this stage, given the wide range of
possibilities consistent with current experimental data.  Particular
parametrizations may emerge as potentially compelling from the
standpoint of flavor theory.  Improved data, particularly for the CHOOZ
angle, will certainly be invaluable in narrowing the range of possible
parametrizations.

\begin{table}[bf]
\begin{center} \begin{tabular}{|c|c|c|c|} 
\hline
& & &   \\
Double Shifts & Right & Left & Middle \\
& & &   \\
\hline
& & &   \\
${\cal V}_{12}\sim a\lambda$& $\theta_{\odot}=\eta_{\odot}+a\lambda$& 
$\theta_{\odot}=\eta_{\odot}+a\cos\eta_{\oplus}\lambda$& 
$\theta_{\odot}=\eta_{\odot}+a\lambda$\\ ${\cal V}_{23}\sim a^\prime 
\lambda$ & 
$\theta_{\oplus}=\eta_{\oplus}+a^\prime\lambda\cos\eta_{\odot}$& 
$\theta_{\oplus}=\eta_{\oplus}+a^\prime\lambda$& 
$\theta_{\oplus}=\eta_{\oplus}+a^\prime\lambda$\\ 
&$\theta_{13}=a^\prime\sin\eta_{\odot}\lambda$
&$\theta_{13}=a\sin\eta_{\oplus}\lambda$
&$\theta_{13}=0$\\
& & &   \\
${\cal V}_{12}\sim a\lambda$& $\theta_{\odot}=\eta_{\odot}+a\lambda$& 
$\theta_{\odot}=\eta_{\odot}+(a\cos\eta_{\oplus}- 
a^\prime\sin\eta_{\oplus})\lambda$& 
$\theta_{\odot}=\eta_{\odot}+a\lambda$\\ ${\cal V}_{13}\sim a^\prime 
\lambda$ & 
$\theta_{\oplus}=\eta_{\oplus}-a^\prime\lambda\cos\eta_{\odot}$& 
$\theta_{\oplus}=\eta_{\oplus}$& 
$\theta_{\oplus}=\eta_{\oplus}$\\ 
&$\theta_{13}=a^\prime\cos\eta_{\odot}\lambda$
&$\theta_{13}=(a\sin\eta_{\oplus}+a^\prime\cos\eta_{\oplus})\lambda$
&$\theta_{13}=a^\prime\lambda$\\
& & &   \\
${\cal V}_{13}\sim a\lambda$& $\theta_{\odot}=\eta_{\odot}$&
$\theta_{\odot}=\eta_{\odot}-a\sin\eta_{\oplus}\lambda$& 
$\theta_{\odot}=\eta_{\odot}$\\ ${\cal V}_{23}\sim a^\prime 
\lambda$ & 
$\theta_{\oplus}=\eta_{\oplus}+(a^\prime\cos\eta_{\odot}-a\sin\eta_{\odot}) 
\lambda$& 
$\theta_{\oplus}=\eta_{\oplus}+a^\prime\lambda$& 
$\theta_{\oplus}=\eta_{\oplus}+a^\prime\lambda$\\ 
&$\theta_{13}=(a\cos\eta_{\odot}+a^\prime\sin\eta_{\odot})\lambda$
&$\theta_{13}=a\cos\eta_{\oplus}\lambda$
&$\theta_{13}=a\lambda$\\
& & &   \\
\hline
& & &   \\
Triple Shifts &Right &Left &Middle\\
& & &   \\
\hline
& & &   \\
${\cal V}_{12}\sim a\lambda$ 
& $\theta_{\odot}=\eta_{\odot}+a\lambda$& 
$\theta_{\odot}=\eta_{\odot}+(a\cos\eta_{\oplus}- 
a^{\prime\prime}\sin\eta_{\oplus})\lambda$ 
&$\theta_{\odot}=\eta_{\odot}+a\lambda$\\
${\cal V}_{23}\sim a^\prime \lambda$
& 
$\theta_{\oplus}=\eta_{\oplus}+(a^\prime\cos\eta_{\odot}- 
a^{\prime\prime}\sin\eta_{\odot})\lambda$& 
$\theta_{\oplus}=\eta_{\oplus}+a^\prime\lambda$&
$\theta_{\oplus}=\eta_{\oplus}+a^\prime\lambda$\\
${\cal V}_{13}\sim a^{\prime \prime} \lambda$ 
&$\theta_{13}=(a^{\prime\prime}\cos\eta_{\odot}- 
a^\prime\sin\eta_{\odot})\lambda$& 
$\theta_{13}=(a^{\prime \prime}\cos\eta_{\oplus}+a\sin\eta_{\oplus})\lambda$& 
$\theta_{13}=a^{\prime \prime}\lambda$\\
& & &   \\
\hline
\end{tabular}
\end{center}
\caption{\label{doubletripleshifts} Leading order shifts in the MNSP 
mixing angles for double and triple shifts ($a,\ a^\prime, 
\ a^{\prime\prime}\sim {\cal O}(1)$).}
\end{table}

\section{CP violation}
Classifying the parametrizations to leading order in $\lambda$ is not 
sufficient for addressing CP violation, since the JDGW invariants are 
given by the product of all the entries of ${\cal U}_{MNSP}$.  Within each 
of the basic classes of models, examples can be constructed by specifying 
the details of the subleading perturbations.  
 To illustrate these points, let us consider 
two representative examples:
\begin{itemize}
\item Single shifts with Cabibbo-shifted atmospheric angles, 
in which ${\cal V}(\lambda)$ is given by
\be
\label{cpexample1}
\mathcal{V}(\lambda)\sim \pmatrix{1& \mathcal{O}(\lambda^3) &  
\mathcal{O}(\lambda^2)\cr
\mathcal{O}(\lambda^3)& 1&a\lambda \cr 
\mathcal{O}(\lambda^2)&-a\lambda& 1}.      
\ee  
\item Double shifts with Cabibbo-shifted atmospheric and solar angles, 
in which ${\cal V}(\lambda)$ is given by
\be
\label{cpexample2}
\mathcal{V}(\lambda)\sim \pmatrix{1& a\lambda &
\mathcal{O}(\lambda^2)\cr
-a \lambda& 1&a^\prime\lambda \cr
\mathcal{O}(\lambda^2)&-a^\prime\lambda& 1}.
\ee
\end{itemize}
To include the effects of CP violation, we allow for a phase $\delta$ of 
{\it a priori} unknown size (though we will assume it is ${\cal O}(1)$) 
that can enter either with the dominant or subleading terms of ${\cal 
V}(\lambda)$. 

The results, which are presented in Table~\ref{jdgwresults}, 
demonstrate explicitly that the CP-violating invariants depend not only on 
the details of the subleading perturbations but also on the type of 
Cabibbo shift, since the mixing angles depend on the particular shift 
scenario.  Unlike the quark sector, the leptonic JDGW invariants 
also depend on whether the CP-violating phase is introduced in the leading 
or subleading perturbations.  

The parametrizations can be classified according to their predictions for 
the JDGW invariants.  As anticipated, the JDGW invariants 
\be
J^{(MNSP)}_{CP}\sim(\lambda-\lambda^4)\sin\delta
\ee 
are much larger than in the quark sector (which is $\sim \lambda^6$). 
JDGW invariants of order $\sim \lambda\sin\delta$ are more common for 
double (and triple) shift scenarios, though single shift models can also 
predict such large effects. 
  
One generic and novel feature of these parametrizations is that the 
CP-violating invariants can be much smaller than naive expectations based 
on the size of the lepton mixing angles, because the effective MNSP phase 
can be additionally suppressed even for $\delta \sim {\cal O}(1)$ (as 
opposed to the CKM phase, which is ${\cal O}(1)$).  We previously 
discussed one example, the class of models based on 
grand unification which satisfy Eq.~(\ref{udagf}),  for which 
the JDGW invariant is ${\cal O}(\lambda^3)$ rather than 
${\cal O}(\lambda)$, as shown in  Eq.~(\ref{ufj}).

Another illustrative example is the single shift scenario of 
Eq.~(\ref{cpexample1}), with right Cabibbo shifts.  In this case, 
$\theta_{13}$ is predicted to be $\sim {\cal O}(\lambda)$ (see 
Eq.~(\ref{atmrshift})), and hence the JDGW invariant is expected to be 
${\cal O}(\lambda)$ if the effective MNSP phase is ${\cal O}(1)$.  As 
shown in Table~\ref{jdgwresults}, this in fact 
occurs if the ${\cal O}(1)$ phase $\delta$ is introduced in the 
dominant $2-3$ mixing, for which 
$J^{(MNSP)}_{CP}\sim\lambda 
\sin\delta$.  However, if $\delta$ is introduced in the most 
hierarchically suppressed perturbations (as in the quark sector), ${\cal 
V}(\lambda)$ is given by 
\be
\label{vex}
\mathcal{V}(\lambda)=\pmatrix{1&(-\frac{a b}{2}+ce^{-i\delta}) \lambda^3 & 
b\lambda^2 \cr
-(\frac{a b}{2}+c e^{i\delta})\lambda^3& 1-\frac{\lambda^2}{2}&a \lambda 
\cr
-b \lambda^2&-a \lambda& 1-\frac{\lambda^2}{2}}\ +\mathcal{O}(\lambda^4)
\ee
($a, \ b,\ c~\sim {\cal O}(1)$).  Right shifts predict a 
suppressed JDGW invariant: $J^{(MNSP)}_{CP}\sim 
\lambda^4\sin\delta$.  In this case, a larger CHOOZ angle does not lead 
to large CP violation because the effective 
MNSP phase is suppressed by ${\cal O}(\lambda^3)$.  

Parametrizations which predict a suppressed effective MNSP 
CP-violating phase  abound in Table~\ref{jdgwresults} and appear to be 
quite generic, reflecting the intriguing possibility that the size 
of the CHOOZ angle is not necessarily correlated with the magnitude of CP 
violation.

\begin{table}[tbf]
\caption{\label{jdgwresults} JDGW invariants for a representative single 
shift scenario with ${\cal V}(\lambda)$ given by 
Eq.~(\ref{cpexample1}) and double shift scenario with ${\cal V}(\lambda)$ 
given by Eq.~(\ref{cpexample2}).  In column 1, the label 
denotes the placement of the phase in the subblock of the perturbing 
matrix. $a$ ($a^\prime$), $b$, and $c$ are ${\cal O}(1)$ parameters 
associated with the ${\cal O}(\lambda)$, ${\cal O}(\lambda^2)$, and ${\cal 
O}(\lambda^3)$ perturbations (see {\it e.g.} 
Eq.~(\ref{vex})). 
} 
\begin{center} \begin{tabular}{|c|c|c|c|} \hline
& & &\\
Single &Right & Left & Middle \\
& & &\\
\hline
& & &\\
12 & $\frac{ab}{16}(\sin3\eta_{\odot}-7\sin\eta_{\odot})
\sin 2\eta_{\oplus}\lambda^4 \sin{\delta}$&
$-\frac{c}{4}\sin 2\eta_{\odot}\sin\eta_{\oplus}\sin 
2\eta_{\oplus}\lambda^3 \sin{\delta}$
&$\frac{ac}{8}\sin 2\eta_{\odot}\sin 2\eta_{\oplus}\lambda^4 
\sin{\delta}$\\
13 & $\frac{b}{4}\cos\eta_{\odot}\sin 2\eta_{\odot}\sin 2\eta_{\oplus} 
\lambda^2\sin{\delta}$&
$\frac{b}{4}\cos\eta_{\oplus}\sin 2\eta_{\odot}\sin 2\eta_{\oplus}
\lambda^2\sin{\delta}$&$\frac{b}{4}\sin 2\eta_{\odot}\sin 2\eta_{\oplus} 
\lambda^2\sin\delta$\\
23 & $\frac{a}{4} \sin\eta_{\odot} \sin 2\eta_{\odot}\sin 2\eta_{\oplus} 
\lambda\sin{\delta}$&
$-\frac{ab}{16}(7\cos\eta_{\oplus}+\cos 3\eta_{\oplus})\sin 2\eta_{\odot} 
\lambda^3 \sin{\delta}$&$
-\frac{ab}{2}\cos 2\eta_{\oplus}\sin 2\eta_{\odot} 
\lambda^3\sin{\delta}$\\
& & &\\
\hline
& & &\\
Double  &Right & Left & Middle \\
& & &\\
\hline
& & &\\
12 & $\frac{aa^\prime}{16}(\sin3\eta_{\odot}
-7\sin\eta_{\odot})\sin 2\eta_{\oplus}\lambda^2 \sin{\delta}$&
$\frac{a}{4}\sin 2\eta_{\odot}\sin\eta_{\oplus}\sin 
2\eta_{\oplus}\lambda \sin{\delta}$
&$\frac{aa^\prime}{8}\sin 2\eta_{\odot}\sin 2\eta_{\oplus}\lambda^2 
\sin{\delta}$\\
13 & $\frac{b}{4}\cos\eta_{\odot}\sin 2\eta_{\odot}\sin 2\eta_{\oplus} 
\lambda^2\sin{\delta}$&
$\frac{b}{4}\cos\eta_{\oplus}\sin 2\eta_{\odot}\sin 2\eta_{\oplus}
\lambda^2\sin{\delta}$&$\frac{b}{4}\sin 2\eta_{\odot}\sin 2\eta_{\oplus} 
\lambda^2\sin\delta$\\
23 & $\frac{a^\prime}{4} \sin\eta_{\odot} \sin 2\eta_{\odot}\sin 
2\eta_{\oplus} \lambda\sin{\delta}$&
$\frac{aa^\prime}{16}(\sin 3\eta_{\oplus}-7\sin\eta_{\oplus}) 
\sin 2\eta_{\odot} \lambda^2 \sin{\delta}$&$
\frac{aa^\prime}{8}\sin 2\eta_{\oplus}\sin 2\eta_{\odot} 
\lambda^2\sin{\delta}$\\
& & &\\
\hline
\end{tabular}
\end{center}
\end{table}

\section{Conclusions}
We are beginning to read the new lepton data, but there is much work to do
before a credible theory of flavor is proposed. In the meantime, we have 
found it illustrative to examine the lepton sector through the lens of 
quark-lepton unification, and investigate parametrizations of the MNSP 
matrix which include Cabibbo-sized perturbations. These 
Wolfenstein-like parametrizations have several novel features, including 
the generic possibility that the size of the CHOOZ angle is not 
necessarily correlated with the observability of CP violation. 

Our approach emphasizes the need for precision measurements of the MNSP
matrix, as the present data is not sufficient to single out a particular
parametrization.  Should the limit of zero Cabibbo mixing prove to be 
meaningful for theory, with improved data we may be able to see flavor 
patterns through the Cabibbo haze.

\section*{Acknowledgments}
This work is supported by the United States Department of Energy under
grant DE-FG02-97ER41209.  A.D. is also supported by the US Department of 
Energy and Michigan Center for Theoretical Physics.

\end{document}